\documentclass{article}

\usepackage[final]{neurips_2022_ml4ps}

\usepackage[utf8]{inputenc} 
\usepackage[T1]{fontenc}    

\usepackage{url}            
\usepackage{booktabs}       
\usepackage{amsfonts}       
\usepackage{nicefrac}       
\usepackage{microtype}      
\usepackage{xcolor}         
\usepackage{amsmath}
\usepackage{amssymb}
\usepackage{graphicx}

\usepackage{wrapfig}

\usepackage{hyperref}       
\begin{document}

\title{Machine-learned climate model corrections from a global storm-resolving model}

%

\author{Anna Kwa \\
  Allen Institute for Artificial Intelligence \\
  \texttt{annak@allenai.org} \\
  \And
  Spencer K. Clark \\
  Allen Institute for Artificial Intelligence, Geophysical Fluid Dynamics Laboratory, NOAA \\
  \texttt{spencerc@allenai.org} \\
  \And
  Brian Henn \\
  Allen Institute for Artificial Intelligence \\
  \texttt{brianhenn@allenai.org} \\
   \And
  Noah D. Brenowitz \\
  NVIDIA \\
  \texttt{nbrenowitz@nvidia.com} \\ 
     \And
  Jeremy McGibbon \\
  Allen Institute for Artificial Intelligence \\
  \texttt{jeremym@allenai.org} \\ 
    \And
  W. Andre Perkins \\
  Allen Institute for Artificial Intelligence \\
  \texttt{andrep@allenai.org} \\ 
    \And
     Oliver Watt-Meyer \\
  Allen Institute for Artificial Intelligence \\
  \texttt{oliverwm@allenai.org} \\ 
     \And
  Lucas Harris \\
  Geophysical Fluid Dynamics Laboratory, NOAA\\
  \texttt{lucas.harris@noaa.gov} \\ 
      \And
  Christopher S. Bretherton \\
  Allen Institute for Artificial Intelligence \\
  \texttt{christopherb@allenai.org} \\ 
}

\maketitle

\begin{abstract}
Due to computational constraints, running global climate models (GCMs) for many years requires a lower spatial grid resolution (${\gtrsim}50$ km) than is optimal for accurately resolving important physical processes. Such processes are approximated in GCMs via subgrid parameterizations, which contribute significantly to the uncertainty in GCM predictions. 
One approach to improving the accuracy of a coarse-grid global climate model is to add machine-learned state-dependent corrections at each simulation timestep, such that the climate model evolves more like a high-resolution global storm-resolving model (GSRM).  
We train neural networks to learn the state-dependent temperature, humidity, and radiative flux corrections needed to nudge a 200~km coarse-grid climate model to the evolution of a 3~km fine-grid GSRM.
When these corrective ML models are coupled to a year-long coarse-grid climate simulation, the time-mean spatial pattern errors are reduced by 6-25\% for land surface temperature and 9-25\% for land surface precipitation with respect to a no-ML baseline simulation. The ML-corrected simulations develop other biases in climate and circulation that differ from, but have comparable amplitude to, the baseline simulation.
\end{abstract}

\section{Introduction}

A recent vein of research uses machine learning to try improve the predictive accuracy of global climate models (GCMs).
GCMs are run at coarse-grid horizontal resolutions of ${\gtrsim}50$ km in order to be tractable for simulations of decades or more.
Cumulonimbus convection and airflow over orography, coastlines and other land-surface heterogeneities are described by length scales smaller than this grid size and are thus approximated by subgrid parameterizations in GCMs. Subjective choices made within these parameterizations contribute significantly to uncertainties in GCM predictions of precipitation, cloud cover, etc. \cite{Shepherd2014}.
On the other hand, global storm-resolving models (GSRMs) run at high spatial (``fine-grid'') resolutions of $<5$ km are able to resolve these processes, but are too computationally intensive to be run for simulations longer than a year or two.

One way to improve the accuracy of GCMs while still maintaining the computational speedup from running at coarse resolution is to train a machine learning (ML) model which is applied ``online'' to update the GCM state at each simulation timestep, with the goal of making the GCM state evolve more like the coarsened state of an equivalent fine-grid GSRM model. 
The evolution of the atmospheric state $\mathbf{x}$ within vertical coarse-grid column $i$ in our ML-corrected GCM is described by $\frac{d\mathbf{x}_i}{dt} = f_i (\mathbf{x}, t) + g(\mathbf{x}_i, \theta)$.
$f_i$ depends on the global state $\mathbf{x}$ and includes the effects of non-local atmospheric circulation as well as the column-local subgrid parameterizations. $g$ is a column-local ML correction depending the column state $\mathbf{x}_i$ and learned parameters $\theta$. 

Previous works have learned ML corrections to the GCM subgrid parameterizations, which were then predicted and applied at runtime in a coarse-resolution simulation with realistic topography \cite{BrethertonEtAl2022, WattMeyerEtAl2021, ClarkEtAl2022}.
Here, we utilize the ``nudge-to-fine'' approach where our reference target is a high-resolution GSRM with 3 km grid resolution.
\cite{BrethertonEtAl2022} previously demonstrated that ML models trained in this manner using a 40-day GSRM simulation improve coarse models' weather forecast skill, land precipitation and surface temperature errors, and diurnal cycle of precipitation.
The novelty of the work lies the training and evaluation of the corrective ML over the entire annual cycle, allowing us to test its performance over seasonal and multiyear timescales; this advance was enabled by the use of a new year-long simulation dataset.
We focus on land precipitation and surface temperature as the top-line metrics by which we gauge improvement over a free-running baseline coarse-grid GCM.

\section{Methods}
\subsection{Data}

\paragraph{Coarse-grid model}
Our coarse-grid model is NOAA's (National Oceanic and Atmospheric Administration) FV3GFS (Finite-Volume Cubed-Sphere Global Forecast System) global atmospheric model \cite{ZhouEtAl2019} run at C48 (${\sim}200$ km) resolution with 79 vertical model levels.
Coarse simulations are carried out using a python-wrapped version of FV3GFS \cite{McGibbonEtAl2021}, which allows for easy customization and setup of nudging and integration of ML models.
Time-varying sea surface temperatures (SSTs) and sea ice fraction are prescribed to be identical to those used in the coarsened fine-grid reference.
We perform a free-running year-long simulation initialized from the coarsened fine-grid state on Jan. 19, 2020. This simulation has no ML corrections and is referred to hereafter as the ``baseline'' run.

\paragraph{Fine-grid model} 
Our reference GSRM simulation is made with the X-SHiELD model, a modified version of FV3GFS with a C3072 cubed-sphere grid (${\sim}$3 km spacing ) and the same 79 vertical levels as the coarse-grid model, run on NOAA's GAEA computing system by collaborators at the Geophysical Fluid Dynamics Laboratory \cite{ChengEtAl2022}. The first three months of spin-up are excluded, resulting in a year-long reference dataset spanning the remaining time from Jan.~19, 2020 through Jan.~17, 2021.

\paragraph{Nudged training simulation}
\label{sec:nudged}
We adopt the `nudging' framework to obtain a training dataset for learning corrective ML tendencies \cite{WattMeyerEtAl2021}. We initialize a coarse-grid simulation from the coarsened fine-grid state at some start time and nudge the simulated coarse-grid state at each successive timestep towards the known time-evolving fine-grid state. The nudged fields are the air temperature, specific humidity and horizontal wind components.
The associated nudging tendencies are $\Delta Q_{a} = - ( a^{n} - \bar{a} ) / \tau$
where $a^{n}$ is a prognostic field in the nudged coarse model, $Q_{a}$ is the source of $a$ from the coarse model physics parameterizations, $\bar{a}$ is the coarsened value of that field in the fine-grid data, and $\tau$ is a constant nudging timescale (3 hours in this work). 

Surface precipitation and downwelling radiative fluxes are systematically biased in the nudged coarse-grid model due to its parameterizations producing less cloud and precipitation than the more accurate fine-grid model. Following the approach taken in \cite{BrethertonEtAl2022, ClarkEtAl2022} we prescribe surface downwelling shortwave and longwave fluxes as well as precipitation during the nudged training run in order to avoid associated biases in the land surface model from feeding back into the atmosphere and affecting the temperature and humidity nudging tendencies. 

If we could perfectly learn the nudging tendencies and the surface radiative fluxes as functions of the nudged coarse atmospheric state, we could apply them to a coarse-grid simulation to make it behave exactly like the fine-grid reference simulation.

\subsection{Machine-learned corrections}
\label{sec:ml_model}
To this end, we train two fully connected dense neural networks (NNs) to separately predict i) vertical profiles of air temperature and specific humidity nudging tendencies and ii) column shortwave transmissivity and downward longwave surface flux. We make the assumption (made by most physical parameterizations used in climate models) that the corrections predicted by the neural networks are column-local, i.e.\ that they only depend on the atmospheric state within a single grid column.

The nudged run training dataset is divided into interleaved blocks of two weeks of training data followed by one week of offline test/validation data, with 6-hour blocks discarded in between to prevent consecutive timesteps across sets. 
1367 timesteps are selected out of the available training data.
At C48 grid resolution, each timestep contains 13824 columns.
We subsample to 15\% of the total number of columns in each timestep in order to reduce memory usage and training time. Thus the training dataset consists of ${\sim}2.8 \times 10^{6}$ samples.
The validation and test datasets used in single-timestep evaluation were each 100 randomly sampled timesteps from the available set.

We train four NNs for each set of outputs using different random seeds and also construct an ensemble model out of each set of four NNs which outputs the median prediction of the ensemble members for each field.

Choices of width, depth and learning rate were guided by a hyperparameter sweep in a randomized grid search with hyperband early stopping \cite{ LiEtAl2018}. Both networks have 3 hidden layers of width 419, use a mean absolute error loss with a L2 regularization penalty of $10^{-4}$. 
The input features for both are the cosine of solar zenith angle, surface geopotential, latitude, and the vertical profiles of air temperature and specific humidity. 
The NN trained to predict air temperature and specific humidity tendencies uses a learning rate of $1.4\times10^{-4}$ while the NN trained to predict surface downward longwave radiative flux and column shortwave transmissivity uses a learning rate of $4.9\times10^{-5}$.
The full configuration is detailed in Appendix \ref{app:model_details}.

\section{Results}
Good performance when evaluated `offline' on predictions over a single time step (Appendix \ref{app:offline}) does not guarantee that the ML corrections will be beneficial when applied within GCMs- the ultimate test of skill is in the `online' setting when the NNs are coupled to the coarse-grid FV3GFS model and correct its state at each timestep. In this setting, feedbacks over many timesteps between the ML and the rest of the coarse-grid model can lead to unphysical  drifts in atmospheric state and even numerical instability. We present results from ML-corrected simulations using each of the four randomly seeded pairs of tendency and radiative flux NNs, as well as a simulation that applies the median correction predicted by the four-member ensemble. 
All ML-corrected simulations ran stably for the full simulation length of 360 days.

\paragraph{Improvements in precipitation and surface temperature}
\label{sec:online_improvement}

\begin{table}[htp]
\centering
\caption{
    Comparison of time-mean surface precipitation metrics over various domains for the year-long baseline and ML-corrected coarse-grid simulations. Since we only have four randomly seeded models to test in ML-corrected simulations, the bottom row cites the range between minimum and maximum relative improvement in each metric across the seeds instead of a standard deviation.
}
\label{tab:precip_metrics}
\begin{tabular}{ l *{6}{c} }
\toprule

\hspace*{2cm} &
\multicolumn{3}{c}{Bias [mm/day]} &
\multicolumn{3}{c}{RMSE [mm/day]}  \\
\cmidrule(lr){2-4} \cmidrule(lr){5-7}  
&
\makebox[3em]{Global} &
\makebox[3em]{Land} &
\makebox[3em]{Ocean} &
\makebox[3em]{Global} &
\makebox[3em]{Land} &
\makebox[3em]{Ocean} \\
\midrule
Baseline & 0.13 & -0.45 & 0.40 & 1.91 & 1.71 & 2.05 \\
\midrule
ML-corrected & 
    \begin{tabular}{@{}r@{}} -0.05 -- \\ 0.04 \end{tabular} &
    \begin{tabular}{@{}r@{}} 0.05 -- \\ 0.26 \end{tabular} &
    \begin{tabular}{@{}r@{}} -0.07 -- \\ -0.01 \end{tabular} &
    \begin{tabular}{@{}r@{}} 1.51 --\\  1.67 \end{tabular} &
    \begin{tabular}{@{}r@{}} 1.29 -- \\  1.55 \end{tabular} &
    \begin{tabular}{@{}r@{}} 1.64 --  \\  1.79 \end{tabular}     \\
\bottomrule
\end{tabular}
\end{table}

\begin{figure}
\centering
  \vspace{-0.5cm}

\includegraphics[width=\textwidth]{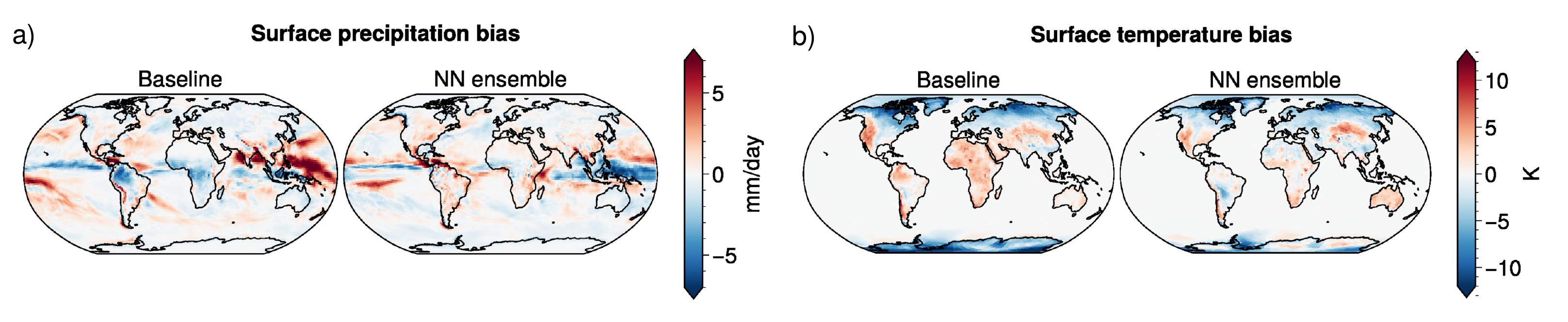} 
\caption{Time-mean error pattern maps of surface precipitation (a) and surface temperature (b) relative to the fine-grid model. Surface temperature is only shown over land and sea ice, as the sea surface temperature is prescribed from the fine-grid reference.}
\label{fig:precip_tsfc_biases}
\end{figure}

Table \ref{tab:precip_metrics} lists the bias and RMSE of time-mean surface precipitation in the baseline and ML-corrected simulations with respect to the fine-grid reference, averaged over land, ocean, and the whole globe.  All ML-corrected simulations significantly improve on the baseline ($13-21\%$ global mean decrease in RMSE).
Figure \ref{fig:precip_tsfc_biases}a shows maps of the time-mean error pattern in precipitation. For conciseness we only show the ML-corrected simulation using the NN ensemble; results are qualitatively similar across the randomly seeded NNs.
Improvements in seasonal-mean land precipitation error are robust across all seasons in all ML-corrected simulations (not shown, available in Appendix \ref{app:seasonal}).

Land surface temperature errors are also reduced by $6-26\%$ in the ML-corrected runs. Figure \ref{fig:precip_tsfc_biases}b shows the  annual-average land surface temperature pattern error in the baseline and NN ensemble runs. The ML-corrected runs reduce a warm bias across Africa and the western United States. While ML-corrected runs  consistently improve land surface temperature from April through September, their behavior in boreal winter is less consistent across random seeds, with NNs generated from two seeds performing comparably or worse than the baseline in these months (Appendix \ref{app:seasonal}).

\begin{wrapfigure}{R}{0.45\textwidth}
  \begin{center}
  \vspace{-1.cm}

\includegraphics[width=0.45\textwidth]{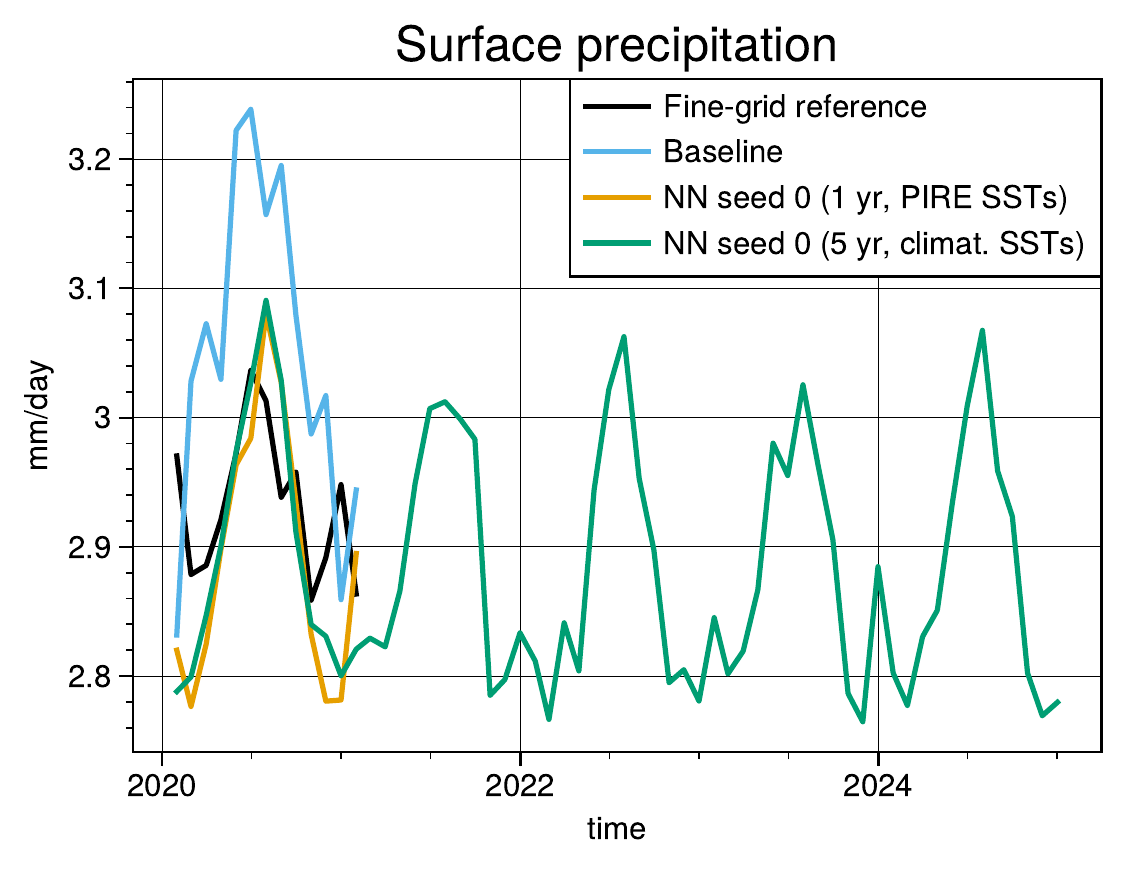} 

\end{center}
\caption{Global- and monthly-mean time series of surface precipitation from the fine-grid, baseline, and ML-corrected simulations. Unlike the yearlong baseline and ML runs, the 5 year ML run uses climatological SSTs and is thus not limited to the time range of the reference run.}
\vspace{-0.6cm}
\label{fig:multiyear}
\end{wrapfigure}

\paragraph{Stability over multiple years}
Would the climate in ML-corrected runs continue to drift, with larger biases developing over multiple years of simulation, or would its biases be largely repeatable in subsequent annual cycles, as in the no-ML baseline? This interannual stability is a prerequisite for use in climate-length simulations. To address this question, we ran a 5-year coarse-grid simulation using the NN with the lowest surface temperature and precipitation errors (seed 0), now forcing it with a climatological annual cycle of SST rather than the date-specific SSTs used in the previous simulations.
Other prognostic runs so far used the SSTs from the yearlong reference X-SHiELD simulation, which did not span the full length of the extended 5 year simulation and did not exactly repeat at the end of the annual cycle. To enable a smoothly-forced 5-year simulation, we instead used a climatological annual cycle of SSTs.
Figure \ref{fig:multiyear} shows time series of global-mean surface precipitation in this extended ML-corrected run as well as the yearlong fine-grid reference, baseline, and ML-corrected runs. As desired, the multiyear ML-corrected run maintains a consistent seasonal cycle of precipitation that matches the fine-grid X-SHiELD reference simulation as well as the 1-year simulation using the same seed and date-specific SSTs (and much better than does the baseline simulation of precipitation).

\paragraph{Biases in ML-corrected simulations}
Several new biases develop in the year-long ML-corrected runs.
The ML-corrected runs develop a strong dry bias in precipitable water of up to -10 mm over the seasonally shifting GSRM-simulated tropical ocean intertropical convergence zones (ITCZs), as well as a northern summertime dry bias at latitudes northwards of ${\sim}50^{\circ}$. The Northern Hemisphere dry bias is also evident in the baseline run, albeit to a lesser magnitude. The baseline model does not share the dry ITCZ bias, but is too moist in most other parts of the subtropics.
A related bias of the ML-corrected runs is a weakened Hadley circulation that is also shifted southward during Northern spring and summer.

ML corrections also significantly reduce tropical precipitation variability.
While the baseline run produces too much extreme precipitation, the ML-corrected runs produce too little. 

We do not yet have a remedy for these ML-induced climate biases. Adding corrective tendencies for horizontal winds might improve circulation biases, however, we found that including this additional ML correction led to air temperature biases that grew quickly within the first 7 days of simulations and ultimately led to numerical instability.  We are currently exploring the use of online novelty detection to regulate the application of ML corrections and prevent predictions on out-of-sample states from pushing the model farther outside of the training data envelope. 

\section{Conclusions}
In this work, we demonstrate that machine-learned corrections trained using fine-grid simulation data improve the surface temperature and precipitation predictions made by coarse-grid climate models. These improvements are robust across the annual cycle. However, the ML corrections introduce changes in circulation and climate biases that differ from those in the baseline coarse-grid climate model. Further refinement of the ML-corrective models may improves these new biases.

\section{Broader impact}
Successful implementation of machine-learned corrections for would improve the accuracy of climate models. In particular, surface precipitation and temperature forecasts in coarse-grid climate simulations could be significantly improved. This would help societies and policymakers to better plan for a future in a warming climate. Conversely, a potential negative outcome would be if decisions were made based off of simulations where the ML corrections degraded forecast skill. Thus it is of utmost importance for researchers to thoroughly validate their models before deeming them production-ready.

\begin{ack}
We thank the Allen Institute for Artificial Intelligence for supporting this work and NOAA-GFDL for running the 1-year X-SHiELD simulation on which our ML is trained using the Gaea computing system.
We also acknowledge NOAA-GFDL, NOAA-EMC, and the UFS community for making code and software packages publicly available.
\end{ack}

\bibliography{main}

\section*{Checklist}

\begin{enumerate}
\item For all authors...
\begin{enumerate}
  \item Do the main claims made in the abstract and introduction accurately reflect the paper's contributions and scope?
    \answerYes{}
  \item Did you describe the limitations of your work?
    \answerYes{}
  \item Did you discuss any potential negative societal impacts of your work?
    \answerYes{}
  \item Have you read the ethics review guidelines and ensured that your paper conforms to them?
    \answerYes{}
\end{enumerate}

\item If you are including theoretical results...
\begin{enumerate}
  \item Did you state the full set of assumptions of all theoretical results?
    \answerNA{}
        \item Did you include complete proofs of all theoretical results?
    \answerNA{}
\end{enumerate}

\item If you ran experiments...
\begin{enumerate}
  \item Did you include the code, data, and instructions needed to reproduce the main experimental results (either in the supplemental material or as a URL)?
    \answerYes{}
  \item Did you specify all the training details (e.g., data splits, hyperparameters, how they were chosen)?
    \answerYes{}
        \item Did you report error bars (e.g., with respect to the random seed after running experiments multiple times)?
    \answerYes{Given as min-max range since N is small.}
        \item Did you include the total amount of compute and the type of resources used (e.g., type of GPUs, internal cluster, or cloud provider)?
    \answerNo{}
\end{enumerate}

\item If you are using existing assets (e.g., code, data, models) or curating/releasing new assets...
\begin{enumerate}
  \item If your work uses existing assets, did you cite the creators?
    \answerYes{}
  \item Did you mention the license of the assets?
    \answerNA{}
  \item Did you include any new assets either in the supplemental material or as a URL?
    \answerYes{Given in supplemental material section \ref{app:code}}
  \item Did you discuss whether and how consent was obtained from people whose data you're using/curating?
    \answerNA{}
  \item Did you discuss whether the data you are using/curating contains personally identifiable information or offensive content?
    \answerNA{}
\end{enumerate}

\item If you used crowdsourcing or conducted research with human subjects...
\begin{enumerate}
  \item Did you include the full text of instructions given to participants and screenshots, if applicable?
    \answerNA{}
  \item Did you describe any potential participant risks, with links to Institutional Review Board (IRB) approvals, if applicable?
     \answerNA{}
  \item Did you include the estimated hourly wage paid to participants and the total amount spent on participant compensation?
    \answerNA{}
\end{enumerate}

\end{enumerate}


\appendix

\section{Workflow and code}
\label{app:code}
The code and experiment configurations needed to reproduce this work are available at the Github repository {\tt https://bit.ly/3RGZK7M} which is archived at Zenodo ({\tt https://bit.ly/3RH05ay}).

\section{Additional ML corrective model details}
\label{app:model_details}

\paragraph{Temperature and humidity tendencies} 
In addition to the hyperparameters and features listed in Section \ref{sec:ml_model}, the following configuration applies to the NNs that predicted temperature and humidity tendencies. 
\begin{itemize}
    \item ML-predicted tendencies of heating and humidity are limited to magnitudes less than 0.002~K/s and $1.5\times10^{-6}$ kg/kg/s, respectively. These limiters are applied as a layer within the dense NN such that the limited outputs are used during optimization. These ranges comfortably extend beyond the nudging tendency minima and maxima in the training data by a factor of 3, but prevent the NNs from making extreme predictions when undesirable feedback between the coarse-grid model and ML corrections lead to atmospheric input states outside the envelope of the training data.
    \item Following the approach of \cite{ClarkEtAl2022}, we exclude (`clip') the uppermost 25 vertical model levels (${\lesssim}150$ hPa) of specific humidity and air temperature state inputs from the feature set. Without doing this, the models' Jacobian matrices showed that the output fields in the boundary layer were as sensitive to inputs from the uppermost 25 model levels as they were to input levels in their immediate locality. \cite{BrenowitzBretherton2019} found that upper level humidity could synchronize with convective processes lower in the atmosphere, leading to NNs learning a backwards causality across these levels that degraded online stability. 
    \item The uppermost 3 vertical levels of temperature and humidity tendency outputs are excluded from the prediction. The ML model always applies a zero corrective tendency for these levels when used online. Differences in the sponge layer damping between FV3GFS and the X-SHiELD reference model lead to large nudging tendencies in these few levels with magnitudes similar to those in the boundary layer, but we do not consider these differences to be part of the coarse model physics that we wish to correct.
\end{itemize}

\paragraph{Surface radiative fluxes}
The NN that predicts surface downward longwave radiative flux and column shortwave transmissivity is used to correct for the effect of systematic cloud biases on the land surface in the coarse run (Section \ref{sec:nudged}). Transmissivity is set to zero in the training data for nighttime columns with zero solar insolation. The predicted column transmissivity is multiplied by the top-of-atmosphere downward shortwave flux to infer the predicted downward shortwave flux at the surface. The downward shortwave and longwave fluxes are applied as state updates in the ML-corrected simulations.

Longwave flux outputs are enforced to be positive or zero, and transmissivity outputs are limited to the range [0, 1]. As in the tendency NN, these limits are applied as a dense network layer such that the limited outputs used in the loss function. 

\begin{figure}[!htb]
\centering
\includegraphics[width=0.45\textwidth]{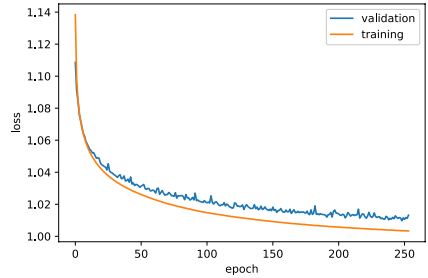} 
\caption{Loss curve from hyperparameter sweep.}
\label{fig:sweep_loss_curve}
\end{figure}

\paragraph{Training and validation loss curves}
Figure \ref{fig:sweep_loss_curve} presents the loss curves for training and validation from the sweep which guided the selection of dense network hyperparameters.

\section{Offline skill}
\label{app:offline}
\begin{figure}[!htb]
\centering
\includegraphics[width=0.75\textwidth]{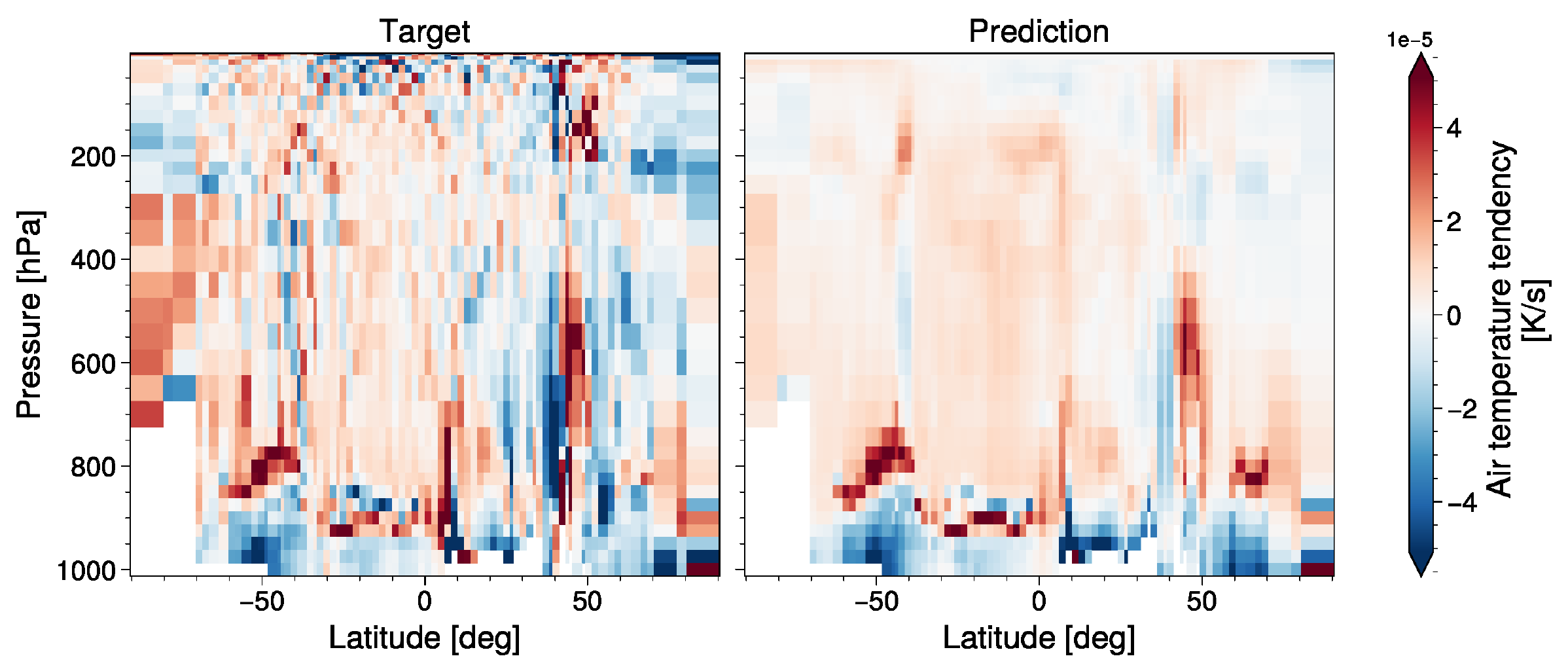} 
\caption{Single-timestep predictions of ML heating tendencies compared to the target values at 2020-02-06 22:30 along a transect at longitude 0$^{\circ}$.}
\label{fig:offline_snapshot_transect}
\end{figure}

``Offline'' skill refers to the NNs' performance in predicting the target correction over a single timestep.
The vertical profiles of the nudging tendencies at a single timestep are quite noisy, so the ML cannot be expected to have perfect skill. Tt is difficult for the NN output to fully describe the variance present in a single timestep of the noisy training data- the ML usually predicts the vertical tendency profile to be much smoother than the the target. This is illustrated in Figure \ref{fig:offline_snapshot_transect} where we show an example of instantaneous target and predicted tendencies along a transect at longitude=0$^{\circ}$. The NN does well to replicate the tendency profiles near the surface, but fails to capture the full magnitude of tendencies higher up in the troposphere. Nevertheless, Figure \ref{fig:offline_snapshot_transect} illustrates that to first order, the ML corrections have learned the desired corrective tendencies over a single timestep.

\begin{figure}[!htb]
    \centering
    \includegraphics[width=\textwidth]{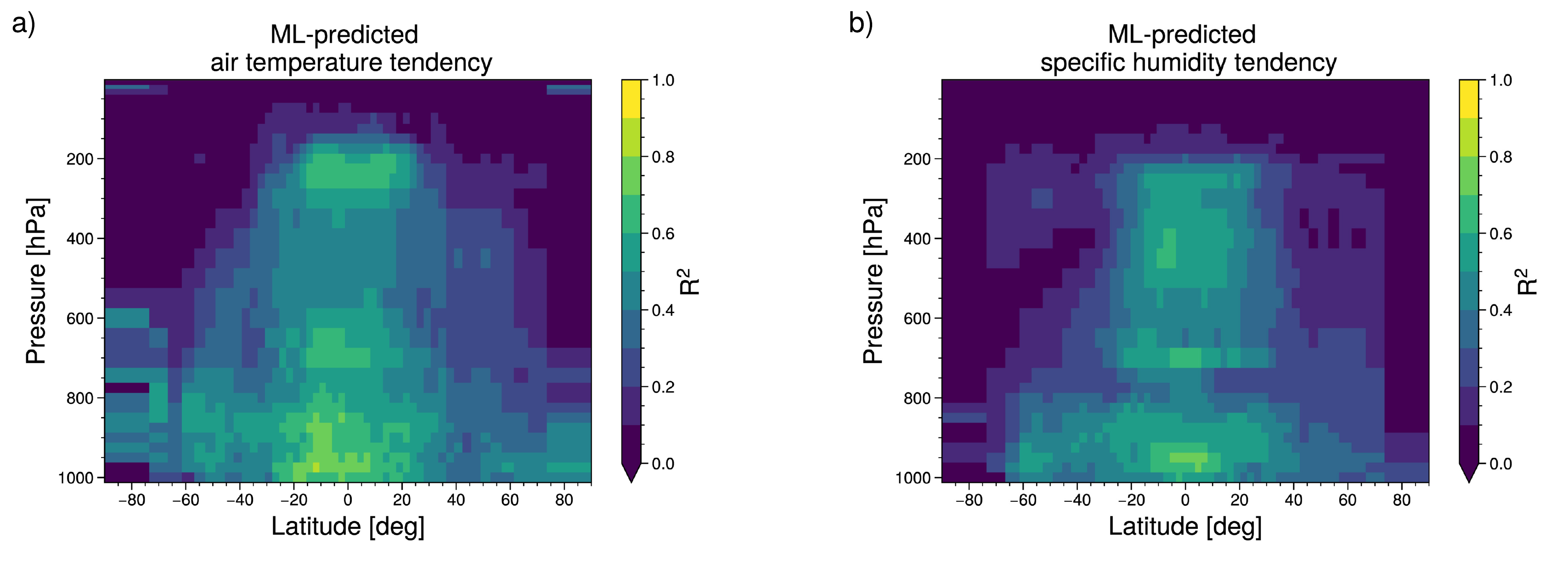} 
    \caption{$R^{2}$ of offline predictions of air temperature and humidity tendencies, averaged over latitude and pressure. Predictions are generated using the NN ensemble; results for individual seeds are similar.}
    \label{fig:offline_R2}
\end{figure}

In Figure \ref{fig:offline_R2} we show the zonal and pressure-level mean coefficient of determination $R^{2}$ on the offline testing data for the ML-predicted corrective tendency fields. Both temperature and humidity tendency predictions are most skillful in the boundary layer and in the tropics, with zonal-mean $R^{2}$ values upwards of 0.8 in the tropical boundary layer and 0.5--0.8 in the tropical free troposphere and extratropical boundary layer. Model skill in the mid-to-upper troposphere degrades to 0.1--0.3 at higher latitudes.

Time-mean offline biases for the downward radiative fluxes are shown in Table \ref{tab:offline_rad_flux}. Their global average biases are small, regardless of neural-net random seed choice: $1.2 \pm 0.7 \mathrm{~W/m}^{2}$ for downward shortwave and $-0.3 \pm 0.2 \mathrm{~W/m}^{2}$ for downward longwave surface radiative flux.

\begin{table}
\caption{Offline metrics for downward surface fluxes. Downward shortwave and longwave fluxes are predicted by the radiative flux NN; total downward flux is the sum of shortwave and longwave predictions.}
\centering
\begin{tabular}{lrrr}
\hline

Surface radiative flux field &  RMSE [W/m$^2$] &  Bias [W/m$^2$] &     R$^2$ \\
\hline
Downward shortwave &        3.8 &        1.2 &  0.99 \\
Downward longwave &        1.2 &       -0.3 &  0.99 \\
Total downward &        3.4 &        0.9 &  0.99 \\
\hline
\end{tabular}

\label{tab:offline_rad_flux}
\end{table}

\section{Performance on seasonal timescales}
\label{app:seasonal}

\begin{figure}
\centering
\includegraphics[width=\textwidth]{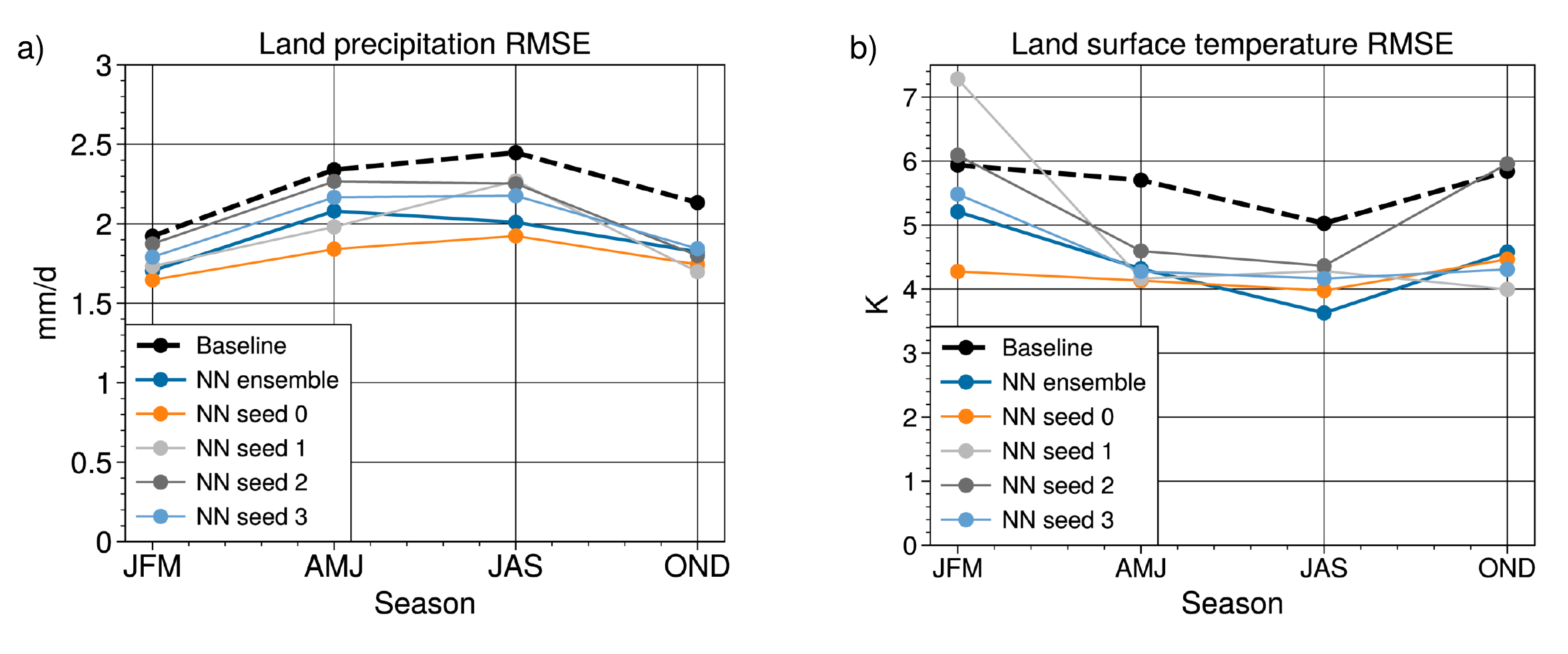} 
\caption{RMSEs of seasonally-averaged surface precipitation (a) and surface temperature (b) in the baseline and ML-corrected coarse-grid simulations.}
\label{fig:seasonal_rmse}
\end{figure}

Figure \ref{fig:seasonal_rmse} plots the seasonally-averaged RMSEs of precipitation and temperature.
All ML-corrected models perform better than the baseline coarse-grid model in seasonal precipitation. 
Seasonal surface temperature improvements are robust across ML models in boreal spring and summer, but two randomly seeded models are on-par with or worse than the baseline during boreal winter. Those two simulations amplify a systematic cold bias during boreal winter at high northern latitudes (${\gtrsim}50^{\circ}$ N) that is also present in the baseline, while the ML-corrected runs with other seeds reduce this bias.

\end{document}